%% file: IEEE_paper_main.tex
\documentclass[%
	final,
	journal, 
    comsoc,
	letterpaper,
	oneside,
	twocolumn,
	nofonttune,%
]{IEEEtran}%
\input{./organization/preamble.tex}%
\input{./organization/makros.tex}%
\input{./organization/settings.tex}%
\makeatletter
\let\blx@rerun@biber\relax
\makeatother
\usepackage{listings}
\usepackage[usenames,dvipsnames]{xcolor}
\usepackage{tikz} \usetikzlibrary{calc, arrows.meta, intersections, patterns, positioning, shapes.misc, fadings, through,decorations.pathreplacing}

\definecolor{ColorOne}{named}{MidnightBlue}
\definecolor{ColorTwo}{named}{Dandelion}
\definecolor{ColorThree}{named}{Plum}

\usepackage{algorithm}
\usepackage{algpseudocode}
\usepackage{tabularx}
\usepackage{newtxmath}

\usepackage{booktabs}
\pgfplotsset{
  grid style = {
   line width = 0.1pt
  }
}
				\newcommand{\disablewr}[1]{#1}%
				\newcommand{\newcommanddisw}[3]{\newcommand{#1}[1]{\disablewr{\textcolor{#2}{#3}}}}%
\renewcommand{\disablewr}[1]{}%
\definecolor{todocol}{named}{red}
\newcommanddisw{\todo}{todocol}{ToDo: #1}%
\definecolor{migucol}{named}{purple}%
\newcommanddisw{\migucom}{migucol}{{@}comment: #1}%
\newcommanddisw{\miguhigh}{migucol}{#1}%
\usepackage{siunitx}
\usepackage{xspace}

\begin{document}%
%
\title{%
 Smart PRACH Jamming: A Serious Threat for 5G Campus Networks
\thanks{This research was supported by the German Federal Ministry for Digital and Transport (BMDV) within the project 5G-CANKRIN under grant number 19OI23012A. The responsibility for this publication lies with the authors. This is a preprint of a work accepted but not yet published at the 2024 IEEE Globecom Workshops (GC Wkshps). Please cite as: J.~R.~Stegmann, M.~ Gundall, and H.D. Schotten: “Smart PRACH Jamming: A Serious Thread for 5G Campus Networks”. In: 2024 IEEE Globecom Workshops (GC Wkshps), IEEE, 2024.}
}
%
\input{./organization/IEEE_authors-long.tex}%
%
%
%
%
%
%
%
\maketitle
%
%
%
%
%
\begin{abstract}%
 Smart jamming attacks on cellular campus networks represent an enormous potential threat, especially in the industrial environment. 
    In complex production processes, the disruption of a single wireless connected \gls{cps} is enough to cause a large-scale failure. 
    In this paper, a smart jamming attack on the \gls{prach} of a 5G system is modeled. This is followed by a practical implementation of the jammer on a testbed based on \gls{oai} and \glspl{softdr}. 
    It is shown that the designed jammer design can interfere a legitimate transmission of a \gls{prach} preamble with a ratio of more than 99.9\%. 
    While less than one percent of the cell resources are interfered compared to broadband jamming. 
    In addition, two different types of jamming signal spectra are compared in relation to their interference capacity.
    The developed attack can be re-implemented based on publicly available source code and \gls{cots} hardware. 
\end{abstract}%
%
%
%
%
%
\IEEEpeerreviewmaketitle
%
%
%
%
%
%
%
%
\tikzstyle{descript} = [text = black,align=center, minimum height=1.8cm, align=center, outer sep=0pt,font = \footnotesize]
\tikzstyle{activity} =[align=center,outer sep=1pt]

\section{Introduction}
\label{sec:intro}
Wireless communications are indispensable for realizing key technologies of \gls{i40}, such as \gls{iot}, digital twins or the increased use of autonomous robots, whereas especially 5G and 6G campus networks will play an essential role. 
A study by McKinsey forecasts B2B 5G \gls{iot} sales at 27 million units in 2025. 
Of these, two-thirds, or 19 million units, will cover distinctive use cases. 
According to the forecast, about half of all distinctive units will be used in \gls{i40} applications in 2030~\cite{mckinsey_5g_2023}.

In many cases, reconfigurable production units can only be effectively controlled wirelessly~\cite{carandbike_team_volkswagen_2021}. 
In contrast to wireline, an inherent protection of the transmission medium from external access cannot be realized. Wireless networks are therefore vulnerable even from outside the company premises. Here, smart jamming attacks pose an enormous threat. 
Compared to classic broadband jamming systems, they can be carried out with less expensive equipment and significantly lower energy consumption. 
However, depending on the type of jammer and victim network, they can cause similarly high damage. 
Especially with some knowledge of the standards and the corresponding signal theory. 
Additionally, smart jammers can often be implemented on a mobile basis, i.e. without a stationary power supply, making them much more difficult to detect and locate \cite{chiarello_jamming_2022}.

\begin{figure}[t!]
\centering
    \includegraphics[width=0.9\columnwidth]{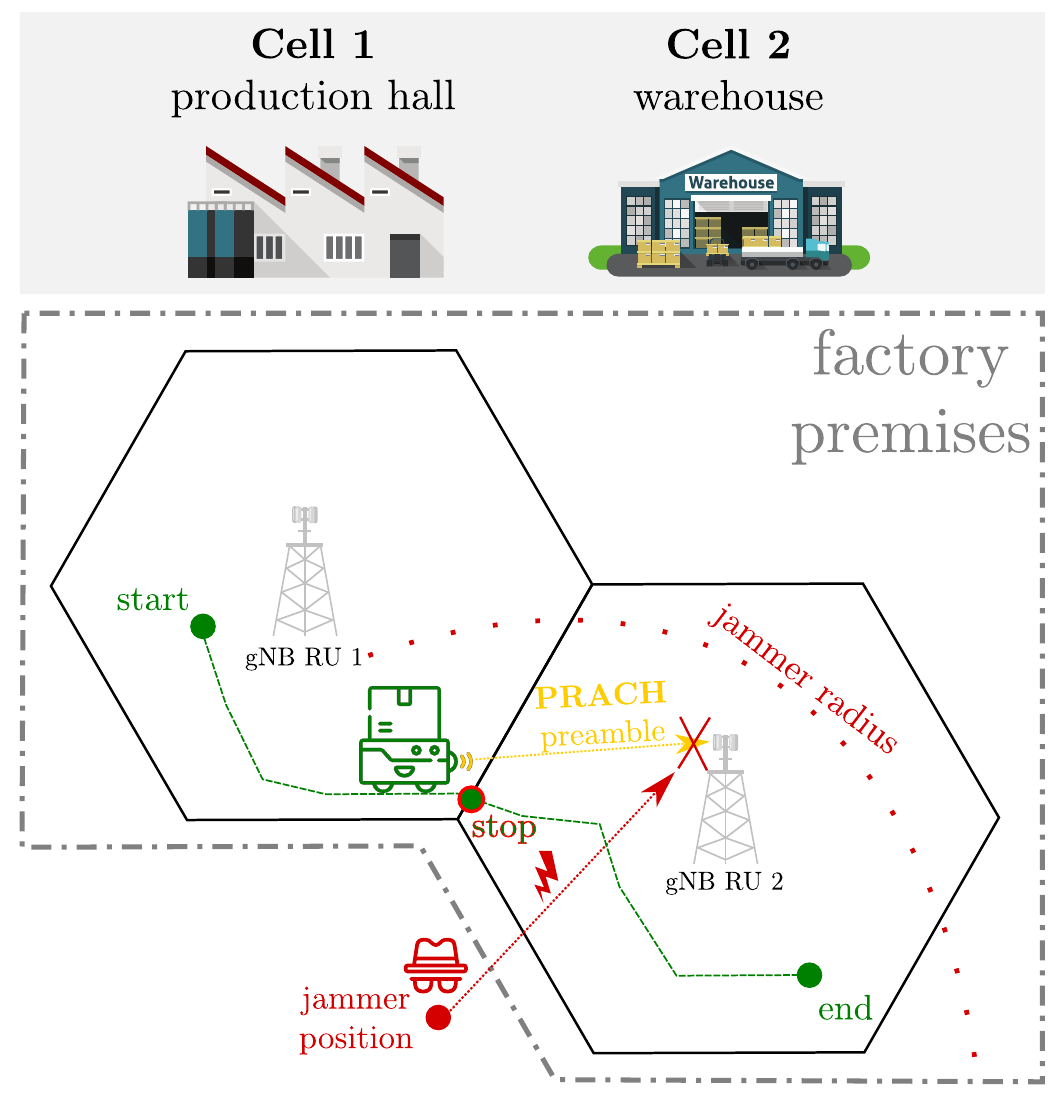}
    \caption{Exemplary trajectory of a wirelessly controlled AGV within a 5G campus network from cell 1 (production) to cell 2 (warehouse).}
\label{fig:prach_attack_factory}
\end{figure}

Particularly interesting are so-called \gls{prach} jammers that aim to prevent the transmission of a preamble from a \gls{uequ} to a \gls{gnodb}. 
A simple example is depicted in \autoref{fig:prach_attack_factory}, where a \gls{agv} is
moving between multiple cells within a factory.
Each time it moves from one cell to another, a handover is triggered. Hence, a preamble is transmitted via the \gls{prach} to the \gls{gnodb} of the next cell. 
Without the transmission of a preamble the connection establishment with the new cell is impossible. 
Due to functional and safety restrictions the \gls{agv} is not able to move without active control connection.  
Since this connection cannot be maintained in the spatial range of the new cell, the \gls{agv} has to stop while leaving the current cell at the latest. Thus, the prolonged failure of the PRACH preamble transmission would automatically result in unintended stop of the its movement.
This can lead to an outage of a production line or even the entire plant. 
A stopped assembly line at large OEMs can quickly cost several thousand USD - per minute \cite{erin_vadala_downtime_2006}. 
Particularly explosive is the combination of a \gls{prach} jammer with high mobility. 
Due to the knowledge of the authors, there are no publicly available \gls{prach} smart jammer implementations using 5G. This is the main motivation for this work. 

Accordingly, the paper is structured as follows:  
The current state of research is presented in Section~\ref{sec:RelWork}. 
Moreover, Section~\ref{sec:smartjamming} compares the effectiveness of known smart jamming attacks.
Subsequently, theory and design of a PRACH smart jammer are introduced in Section~\ref{sec:design of smart PRACH jammer}. This is followed by Section~\ref{sec:testbed}, where the testbed, the implementation, and the results are detailed. Finally, Section~\ref{sec:Concl} concludes the paper. 

\section{Related Work}
\label{sec:RelWork}
This section details the state of research. \textit{Lichtman et al.}~\cite{5g_acia_security_2022} give a overview of the security requirements for 5G systems in industrial environments. 
The reference to the safety standard IEC 62443, where resource availability is one of the foundational requirements, is also made. Jamming attacks usually target IEC 62443 foundational requirements 7 (FR7), i.e. resource availability. 
\textit{Arjoune and Faruque}~\cite{arjoune_smart_2020} summarise key jamming attack vectors on 5G. In particular, there is also a distinction between the different physical channels of 5G. Further, various jamming defence mechanisms are discussed. However, they remain rather technical superficial in both aspects, focus on classical jammer designs and provide little practical evidence for their statements. 

In \cite{pirayesh_jamming_2022}, the vulnerability of wireless networks to jamming attacks is addressed with a more general focus. 
A large part deals with cellular networks, with a clear focus on LTE.
Some of the theoretically discussed findings can also be applied to 5G. As jamming countermeasures, they primarily identify the exploitation of MIMO functionalities, spectrum spreading techniques and dynamic resource allocation. 

\textit{Ludant and Noubir}~\cite{ludant_sigunder_2021} have developed an overshadowing attack that superimposes parts of the 5G \gls{pbch}.
This attack vector exploits the non-existent integrity control of the \gls{pbch} and modify network parameters so that a UE connects to a non-functional malicious cell.

\textit{Chiarello et al.}~\cite{chiarello_jamming_2022} discuss objectives of a jammer in the area of 5G campus networks. The requirements for a resilient infrastructure are also posed on the basis of theoretical considerations.

\section{Smart jamming}
\label{sec:smartjamming}
In contrary to broadband jamming, smart jamming refers to attacks that involve the intelligent use of available resources to interfere legitimate signals in the radio frequency (RF) range. The aim of smart jamming is therefore to cause as much damage as possible with minimal use of resources. This is referred to as effectiveness. 
The reference and counterexample is broadband jamming. 
In this case, the relevant frequency band is interfered over the entire bandwidth with maximum transmission power over the entire time period.
Smart jammers, on the other hand, try to use the shortest possible duty cycle, narrow bandwidth and low power. 

\subsection{Effectiveness Analysis}
\label{subsec:effectiveness}
Jammers can have different objectives depending on the motivation of the attacker \cite{chiarello_jamming_2022}. 
In an \gls{embb} scenario, spectral efficiency mitigation is an objective. Thus, superimposed noise can greatly reduce the number of bits per symbol. 
In \gls{urllc}, on the other hand, increasing the bit or symbol error rate is a promising goal, since this greatly increases the number of data packet re-transmissions. 
This usually leads in such environments to a violation of strict latency and synchronization requirements.
While these two attacks aim to reduce the communication quality (user plane), control plane attacks are also conceivable. In this case, the establishment or sustainment of the connection between \gls{uequ} and \gls{gnodb} is targeted. 
In order to rate attack vectors, the following metrics are introduced.
\subsubsection{Power efficiency}
Numerical ranking on a scale of 1-10 based on the bandwidth capacity for the attack, the relative proportion of resource elements to be jammed in relation to the entire resource grid.
\subsubsection{Effort efficiency}
Ranking on a scale of 1-10 based on the required degree of resource grid synchronisation, a priori knowledge, traceability of the jammer, implementation effort (hardware costs and coding), and the potential damage. 

\subsubsection{Efectiveness}
Combination of power and effort efficiency, whereas a higher value indicates increased effectiveness. 

\begin{figure}[htbp]
    \centering
    \includegraphics[width=0.85\columnwidth]{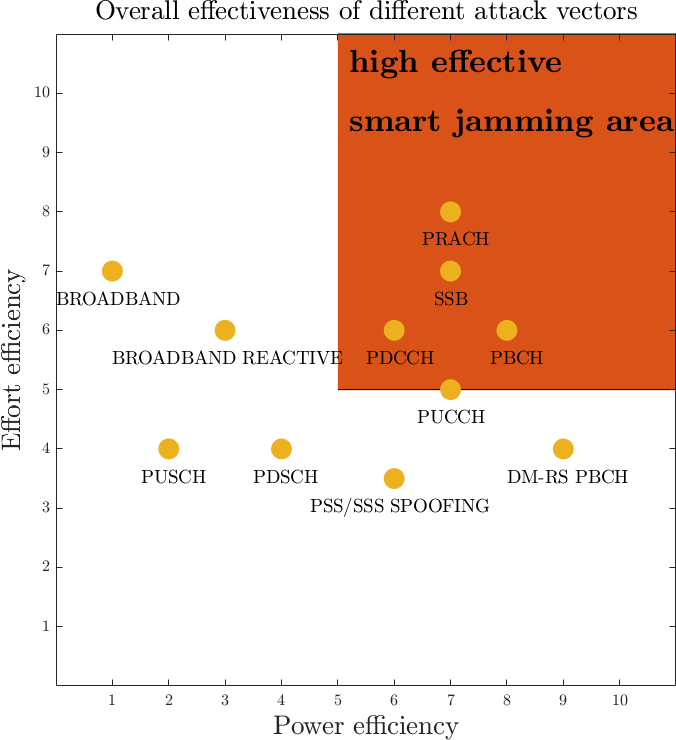}
    \caption{Jamming effectiveness based on effort and power efficiencies.}
    \label{fig:effectiveness}
\end{figure}

The resulting classification can be found in Figure~\ref{fig:effectiveness}. According to objectives within this paper, attacks on the \gls{pbch}, the \gls{prach}, or the control channels in the uplink and downlink are of particular interest (red marked area).
\subsection{Motivation for \gls{prach} Jamming}
\label{subsec:prach_attack}
Besides the already mentioned consequences on handover procedures between cells, the initial entry of new \glspl{uequ} is also prevented. Furthermore, \glspl{uequ} that have fallen out of synchronization with \gls{gnodb} no longer have a chance to reconnect to the cell. The same applies to devices that are not continuously connected to the cell (for example, for energy-conserving reasons). As these scenarios, \glspl{cps} networked as \glspl{uequ} can lead to significant malfunctions up to complete outage of larger assemblies. Hence, the \gls{prach} is a suitable attack vector. 
    Hence, the following situations trigger a \gls{ra} procedere and thus the (re-)transmission of a preamble:
\begin{itemize}%
\item%
    Initial access
\item%
    \gls{rrc} connection reestablishment
\item%
    Handover
\item%
    Non-synchronised RRC\_CONNECTED state
\item
    Transition from RRC\_INACTIVE state
\item
    Time-alignment for secondary cells
\item 
    System Information (SI) requests
\item%
    Beam failure recovery
\end{itemize}%


\section{PRACH jammer design}
\label{sec:design of smart PRACH jammer}
The \gls{pbch} fulfils two essential functions as a DL channel. 
On the one hand, it carries the \gls{mib} with the elementary information for locating and selecting a cell. 
On the other hand, cell synchronisation in time and frequency are enabled by broadcasted beacon signals. 
In the uplink the \gls{prach} provides the basic function to (initially) register in a cell. 
\subsection{Random Access Procedure}
\label{subsec:ra_proc}
In the 5G \gls{ra} procedure, a basic distinction can be made between contention-based and contention-free execution. 
In the contention-free variant, a \gls{prach} signature is uniquely assigned to the \gls{uequ} via the \gls{rrc} before the preamble transmission. 
However, this is only possible if the \gls{uequ} has already established a connection to the network. 
With the contention-based methods, the \gls{uequ} selects a possible signature at random. Since the number of signatures is based on cyclic shifts of \gls{zc} sequences of the same root index and length, their total number is limited. 
Thus, the probability that two \glspl{uequ} transmit the same signature to the \gls{gnodb} at the same time is given.  
This can take place separately after the preamble transmission (4-step \gls{ra}) or already be initiated with it (2-step \gls{ra}). 
For each form of contention and number of steps, a preamble is always transmitted via the \gls{prach}.
\\
\begin{figure}[htbp]
    \centering
    \includegraphics[width=0.8\columnwidth]{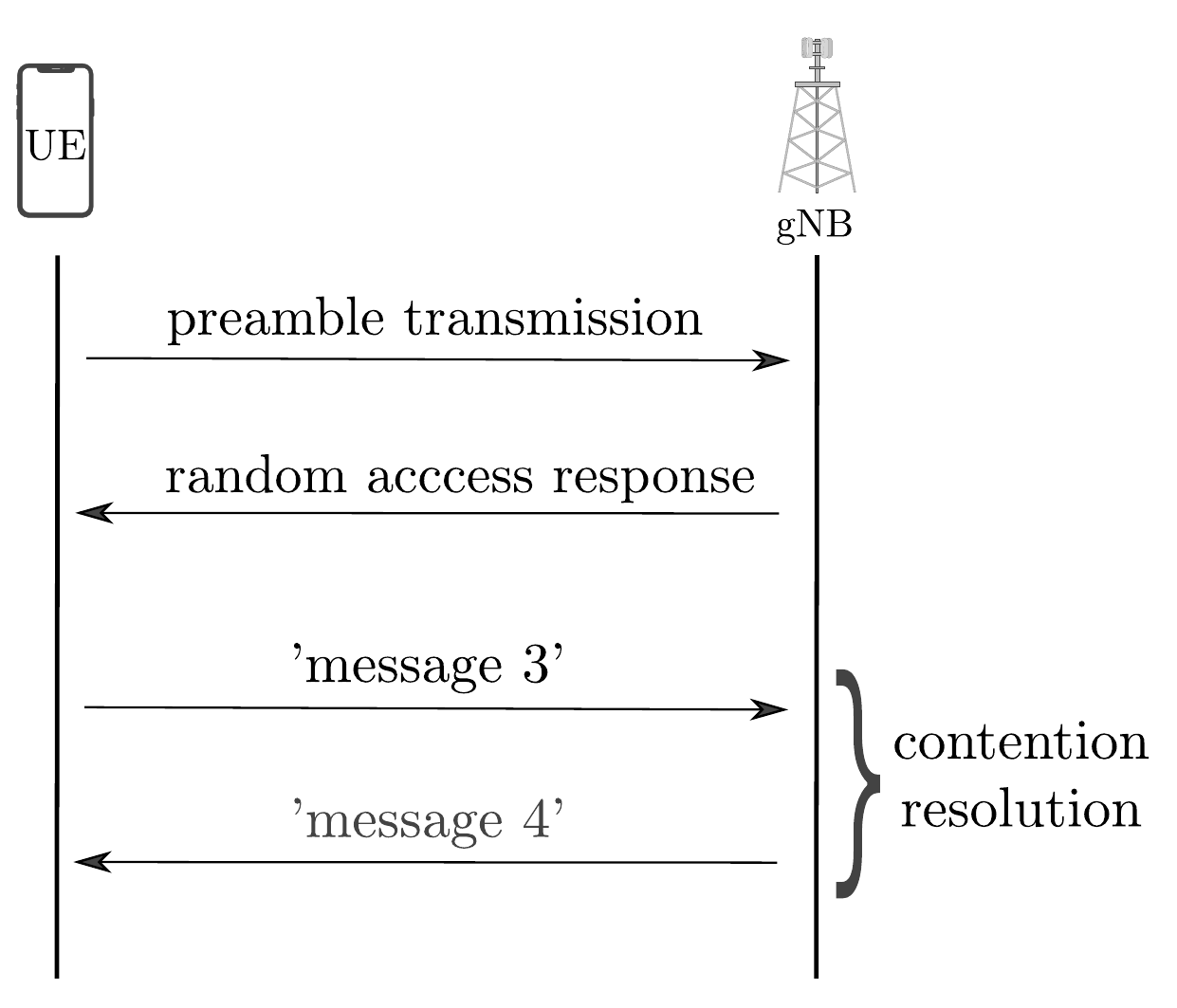}
    \caption{\centering 4-step contention-based Random Access Procedure}
    \label{fig:rach_process}
\end{figure}
As shown in Figure \ref{fig:rach_process}, a preamble 
is transferred from the \gls{uequ} to the \gls{gnodb} via the \gls{prach}. 
If the reception is successful, the \gls{gnodb} sends the \gls{rar}.
This contains a mapping between the received signature and an allocation of a temporary identifier 
and network resources.
If several \glspl{uequ} have sent the same signature at the same time, both UEs assume that the ID and resources have been allocated to them. 
Hence, all \glspl{uequ} send a control message (message 3) to the \gls{gnodb} with an unique identifier. 
The \gls{gnodb} responds with message 4 which unique identifier should be given access to the network. 
All other \glspl{uequ} 
have to repeat the complete \gls{ra} procedure. 
Since a smart jammer that attacks the \gls{prach} impedes the \gls{gnodb} from receiving preambles at all, it prevents the first step of the \gls{ra} procedure. 
To obtain the location of the \gls{prach}, each \gls{uequ} first synchronises with the network via the PSS and SSS 
before decoding the \gls{mib}. Using the \gls{mib}, information about the cell, primarily the location of the coreset within the PDCCH can be identified. 
Based on an indication within the coreset, \gls{sib}1 can be received. 
Using the information from \gls{sib}1, \gls{mib} and a-priori standardisation knowledge, the exact location of the \gls{prach} in the time and frequency domain can be determined. 

\subsection{\gls{prach} Configuration}
\label{subsec:prach_conf}
The possible positioning of the \gls{prach} within the resource grid can vary and depends on various cell parameters. 
The corresponding parameters can be found in Table \ref{tab:prach_channel_configs}. 
\begin{table}[h]
\footnotesize
    \centering
	\caption{\gls{prach} channel configuration based on TS 38.211~\cite{etsi_ts_2018}.}
	\label{tab:prach_channel_configs}
	\begin{tabular}{@{} c c @{}}
		\toprule[1.5pt]
            \textit{Parameter} & Value\\
		\midrule[0.8pt]
            Preamble length $L_{RA}$ & 139 (Short) \\
            \gls{prach} length $M_{PRB}$ in PRBs  & 12 \\
            Number of \gls{prach} occasions $K$  & 1\\
            Frequency Offset & 0\\
            Preamble Format & A2 \\
            \gls{prach} frame numbers $n_{SFN}$ & all odd SFNs \\
            Subframe Number & 9\\
            Slot Number $n_{slot}^{RA}$ (for $\mu = 1$) & 1 \\ 
            Starting Symbol & 0\\
            \gls{prach} slots within a subframe & 1\\
            \gls{prach} occasions per \gls{prach} slot & 3\\
            \gls{prach} duration in symbols & 4\\
            \bottomrule[1.5pt]
	\end{tabular}
\end{table}
Depending on the spatial cell size, a distinction is made between short and long preambles. 
Long preambles are used in spatially large cells with low subscriber density, whereas short preambles are predominantly used in spatially small cells with medium and high subscriber density. 
Long preambles have a sequence length of 839, short ones of 139. 
They occupy the corresponding number of subcarriers (preamble bandwidth $M$) in the frequency range. 
The total occupancy in the frequency range ($K \cdot M$) then depends on the number of frequency occasions $K$.
The frequency offset determines the exact position within the total channel bandwidth.

Depending on the temporal \gls{prach} occasion and preamble duration, a certain number of preambles can be transmitted in each \gls{prach} slot.
The time-domain length of a preamble depends on the so-called preamble format. On the one hand, this determines the temporal length of the cyclic prefix and, on the other hand, the number of temporal repetitions of the same sequence within a preamble. 
In this work, a short preamble ($L_{RA}$~=~139~=~$M$) is assumed because the cells of campus networks can usually be considered as comparably small.
With only one \gls{prach} occasion ($K$~=~1) and an offset of 0 in the frequency domain, the \gls{prach} slot occupies 12 PRBs starting at frequency point $A$ of the cell. 
In the time domain, the distribution is determined by the configuration index of 98 that specifies the preamble format of A2. 
This is a cyclic prefix followed by four repetitions of the sequence in the time domain and results in a length of four \gls{ofdm} symbols per preamble. 
Moreover, each \gls{prach} slot contains three time-domain \gls{prach} occasions. Thus, three preambles can be transmitted in succession. 
In this case, each subframe contains exactly one \gls{prach} slot ($N_{sl}=1$). 
Also,
there is exactly one subframe containing a \gls{prach} slot per frame ($N_{sf}=1$), whereas the \gls{prach} slot is located in the ninth (last) subframe. If the SCS is 30 kHz, the \gls{prach} can  be found in the second slot of this subframe ($n_{slot}^{RA}=1$). 
Since the starting symbol is 0, the first 12 of the 14 symbols in the 20th slot of the frame are assigned to the \gls{prach} ($N_{sy}=12$). 

If $x = 2$ and $y = 1$ for $n_{SFN}\mod x = y$, each frame with an odd \gls{sfn} contains a \gls{prach} slot ($T_{ra}$~=~20~ms). 

\begin{equation}
    \label{eq:rel_ra}
    \rho_{ra} =  \underbrace{\frac{10~\mathrm{ms}}{T_{ra}}}_{\text{\gls{prach} period}} \underbrace{\frac{N_{sf} \cdot N_{sl} \cdot N_{sy}}{10 \cdot 2^{\mu} \cdot 14}}_{\substack{\text{temporal} \\ \text{occupation}}} \underbrace{\frac{2^{\mu} \cdot 15~\mathrm{kHz} \cdot M_{} \cdot K_{}}{B_{cell}}}_{\substack{\text{bandwidth} \\ \text{occupation}}}
\end{equation}
Hence, the \gls{prach} occupies only $\rho_{ra}$ $\approx$ 0.22~\% of the total available resource elements. Thus, if $B_{cell}$ = 40~MHz, a bandwidth of about 4~Mhz is sufficient to jam the \gls{prach}. 
In addition, only a duty cycle of 20~ms and a jamming period of about 0.5~ms are required. 
This increases the effectiveness of the smart jammer. 

\subsection{\acrlong{zc} sequences}
\label{subsec:zadoff-chu sequences}
The \gls{prach} preambles are composed of \gls{zc} sequences that belong to the constant amplitude zero autocorrelation waveforms. 
For an odd sequence length $N$, the \gls{zc} sequence can be expressed by 
\begin{equation}
    \label{eq:zadoff_chu}
    x_{U}(k) = \exp \left(j \frac{\pi U k(k+1)}{N}\right), \quad k = 0,1,\cdots,(N-1)
\end{equation}
Cyclically shifted \gls{zc} sequences of the same root index $U$ have an autocorrelation of zero. 
For a fixed sequence length $N$, the number of root indices $U$ generating unique sequences depends on the number of integers relatively prime to $N$. 
In addition, two sequences of different prime root indices $U_1$ and $U_2$ have a constant cross-correlation of $1/\sqrt{N}$ if the difference $U_2-U_1$ is relatively prime to $N$.
Moreover, if $N$ is itself a prime number, the DFT of a \gls{zc} sequence is also a \gls{zc} sequence of the same length $N$. 
Hence, $N$ is chosen to be a prime number for 5G. 

In practice, orthogonal preambles are used to reduce the interference of neighbouring cells. 
Constant amplitude significantly simplifies processing within the transmit and receive amplifier. 
The cyclic shifts and different root indices allow a unique assignment to the cell.  
However, the uniqueness of a sequence is only maintained with multipath propagation as long as the maximum propagation difference is smaller than the minimum possible cyclic shift in the time domain. 
For this reason, not all possible shifts are used. Instead, depending on the cell size, a certain minimum step size between two shifts is usually used. Due ro the higher sequence length $N$, long preambles are preferred in spatially large cells. 

\subsection{Jamming Spectra and Investigated Quantities}
Two different jamming spectra were investigated. Spectrum 1 (S1) is an Artificial White Gaussian Noise (AWGN) in the time domain with the prescription $\underline{f}_{S1,k}$ in the frequency and $\underline{t}_{S1,k}$ in the time domain.
\begin{align}
    \underline{f}_{S1,k} = \begin{cases}
        A_f + 0 j = const. &, k \in [0 \enspace (L_{RA}-1)]
        \\
        0 + 0j &, k \in [L_{RA} \enspace (N_{DFT}-1)]
    \end{cases} 
\end{align}

\begin{equation}
    \underline{t}_{S1,k} = IFFT \left(\underline{f}_{S1,k}\right) = A_t \cdot randn(k) \enspace,k \in [0 \enspace (N_{DFT}-1)]
\end{equation}

For S1, the power density spectrum is real and constant over all \gls{prach} subcarriers ($k<L_{RA}$). 
Spectrum 2 (S2) is an AWGN in the frequency domain over the affected subcarriers with the prescription:
\begin{align}
    \underline{f}_{S2,k} = \begin{cases}
        A_f \cdot randn(k) &, k \in [0 \enspace (L_{RA}-1)]
        \\
        0 + 0j &, k \in [L_{RA} \enspace (N_{DFT}-1)]
    \end{cases} 
\end{align}
\begin{equation}
    \underline{t}_{S2,k} = IFFT \left(\underline{f}_{S2,k}\right) \enspace ,k \in [0 \enspace (N_{DFT}-1)]
\end{equation}
The $randn(k)$ function generates random numbers that follow the standard complex normal distribution:
\begin{equation}
  n(\underline{z}) = \frac{1}{\pi} e^{-|\underline{z}|^2}
\end{equation}
The amplitude of the interference signal is given by:
\begin{equation}
  A_f = A_n \cdot A_{SNR} = A_N \cdot 10^{\frac{-SNR}{20}}
\end{equation} 

\subsection{Metrics}
\label{subsec:metrics}
In order to evaluate the PRACH jammer, metrics are defined. Thus, $N_{RA,s}$ is the number of successful and $N_e$ the number of invalid intervals. 
An interval is successful if a preamble of the \gls{uequ} is received at the \gls{gnodb}. 
$N_{P,j}$ is the total number of preambles over all intervals contained in the associated measurement series that were successfully jammed. 
$\bar{N}_{P,t}$ indicates the average number of preambles sent per valid interval and is calculated by 
\begin{equation}
  \bar{N}_{P,t} = \frac{N_{P,j} + N_{RA,s}}{N_i-N_e}.
\end{equation}
$E_{P,j}$ gives the ratio between the preambles successfully received by the \gls{gnodb} and the all preambles sent 
\begin{equation}
   E_{P,j} = \frac{N_{RA,s}}{N_{P,j} + N_{RA,s}}.
\end{equation}
This is also a conservative estimate of the probability of a single successful preamble transmission.
The actual probability will most likely be lower, since no further preambles are sent by the \gls{uequ} after \gls{rar} reception.
$E_{s}$ indicates the ratio between the intervals with successful \gls{ra} procedure and all valid intervals:
\begin{equation}
   E_{s} = \frac{N_{RA,s}}{N_i-N_e}
\end{equation}
\section{Testbed and Implementation}
\label{sec:testbed}
This section details the testbed, the implementation, and the results.

\subsection{Testbed}
\begin{figure}[htbp]
\centering
{\includegraphics[width=0.8\columnwidth]{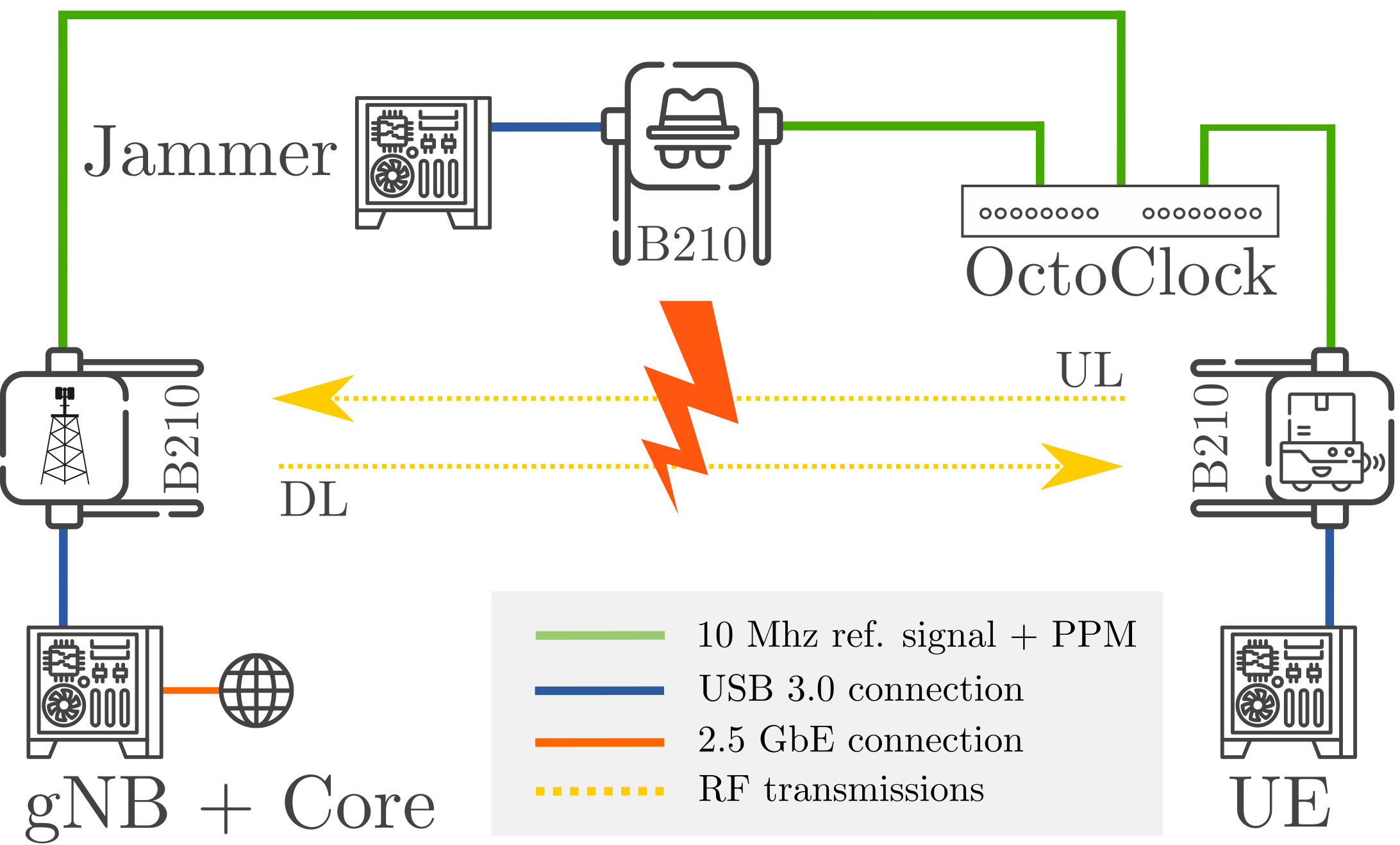}}
\caption{Smart jamming testbed consisting of three tower PCs, SDRs, and a reference clock.}
  \label{fig:testbed}
\end{figure}

The testbed, shown in Figure~\ref{fig:testbed}, 
consists of three SDRs, each one connected to a PC, and a reference clock. The setup can also be operated without latter, but the drifts of the \gls{usrp} clocks distort longer measurement campaigns. One of the PCs is running both gNB and Core, while the other ones are running OAI UE modules. 

Since the entire signal processing is outsourced to the host, the hardware requirements in terms of computing power are comparatively high. 
To meet these requirements, we tower PCs are used, taking \cite{ettus_research_oai_2023} into account.
Further information on the software and hardware used can be found in Table \ref{tab:params_testbed}. To improve performance, a low-latency kernel is used and the sleep modes in the C-States are deactivated. 
\begin{table}[h]
\footnotesize
    \centering
	\caption{Testbed Hard- and Software Specification}
	\label{tab:params_testbed}
	\begin{tabular}{c c c c}
		\toprule[1.5pt]
            & \gls{gnodb} + Core & \gls{uequ} & Jammer \\
		& {\textit{Tower 1}} & {\textit{Tower 2}} & {\textit{Tower 3}}\\
		\midrule[0.8pt]
            SDR & \multicolumn{3}{c}{Ettus Research \acrshort{usrp} B210} \\
            Ref. clock & \multicolumn{3}{c}{Ettus Research OctoClock CDA-2990} \\
            \midrule[0.8pt]
		CPU & i9-12900KS & i9-10900X & i9-12900KS \\
            RAM & 128GB DDR5 & 256GB DDR4 & 64GB DDR5 \\
            \midrule[0.8pt]
            OS & \multicolumn{3}{c}{Ubuntu 18.04.6 LTS 64bit}  \\
            Kernel & \multicolumn{3}{c}{5.16.11-051611-lowlatency} \\
            \midrule[0.8pt]
            \multirow{2}{*}{5GS} & {\gls{oai} 5G RAN +} & \gls{oai} 5G & Customized \\
             & \gls{oai} 5G CN & \gls{uequ} & \gls{oai} 5G \gls{uequ}\\
		\bottomrule[1.5pt]
	\end{tabular}
\end{table}

\subsection{Implementation}
\label{sec:implementation}
This section details the implementation. The victim cell for the jammer was set up taking into account the limitations of \gls{oai} and the B210 USRPs in the 5G \gls{newr} n78 band. This is motivated by the fact that the authorities in Germany, i.e. the Federal Network Agency, have earmarked the frequency range between 3.7 and 3.8 Ghz for campus networks.

To implement the jammer, the \gls{oai} \gls{uequ} softmodem code has been modified. Besides the selection of the cell to be jammed, three essential manipulations were carried out:
\begin{enumerate}%
\item%
	Transmit in each \gls{prach} slot of the cell 
\item%
	Transmit noise in all three \gls{prach} occasions per \gls{prach} slot
\item%
	Transmit some specific noise spectrum instead of a legitimate preamble
\end{enumerate}%
The steps for cell synchronization (\gls{pbch} decoding, \gls{sib}1 reception, determination of the \gls{prach} position) from section \ref{subsec:ra_proc} were fully adopted.  
Therefore, both the required expertise and the overall implementation effort for this jammer type are comparatively low. 
The first manipulation was performed from the point on where a regular \gls{uequ} initiates the \gls{ra} procedure with the transmission of a preamble. 
Instead of transmitting in only one of the slots and  waiting for \gls{rar} for a period of time first, the code was modified to transmit in all \gls{prach} slots of the cell. 
Additionally, the length of the IQ samples to be transmitted was tripled in the time domain to cover all \gls{prach} occasions within the slot. 
In principle, this is also possible with the extension of the transmission spectrum in the frequency domain, but this was not necessary with $K=1$. 
Then, the possibility was created instead of a (repeated) \gls{zc} sequence to send an arbitrary spectrum in the form of IQ samples in the respective slots. In order to reproduce the results, the source code is publicly available on GitHub\footnote{https://github.com/dfki-in-icc/5G-PRACH-Smart-Jammer}. 

Further, $SNR=-$6~dB is chosen for the measurements, leading to $A_{SNR}\approx$ 2. Thus, the amplitude of the jamming signal is about twice that of the legitimate preamble.  
To achieve the expected performance of the jammer, approximately four times the transmit power must be applied compared to the \gls{uequ}. 
As \glspl{uequ} are mobile devices with limited antenna power and the \gls{prach} is an uplink channel, this signal strength can easily be provided by SDRs. Thus, the results obtained are representative for all spatial points where this power ratio is satisfied. 

\subsection{Results}
\label{sec:results}
Both spectra are validated in jamming intervals of 60~s and a total of $N_i=800$ runs. S1 was validated again in a second stage at intervals of $N_i=40$ intervals of 600~s. 
To better classify the results, an additional 800 runs without active jammer were performed.  
\gls{uequ} and jammer were restarted for each interval, whereas the jammer is started first, followed by the \gls{uequ} with a 10-second delay. 
In the specified interval period, the \gls{uequ} starts and then attempts to establish a connection with the \gls{gnodb}. 
After the interval has expired, the \gls{uequ} is terminated and the jammer is terminated with a further 10-second delay.
Thus, the jammer is 20 seconds longer active than the \gls{uequ}. 
This ensures that preambles are always transmitted while the jammer is running. 

Every 10 frames, i.e. 100~ms, the \gls{uequ} tries to transmit a preamble as long as no \gls{rar} has been received from the \gls{gnodb}. 
This leads to a maximum transmission of about 450 preamble in 60~s and $\approx$ 5000 in 600~s. 
As soon as a preamble has been successfully received, the \gls{gnodb} responds with an \gls{rar}. 
The \gls{uequ} terminates the inital access and does not transmit any further preambles in the \gls{prach}. 
For this reason, with the reception of exactly one received preamble, the failure of the jammer and a successful \gls{ra} procedure can be assumed. If no preamble is received, the attack was successful in the corresponding interval. 

\begin{figure*}[htbp]
\centering
 \subfloat[800 intervals of 60s with deactivated jammer.]{\includegraphics[width=0.95\columnwidth]{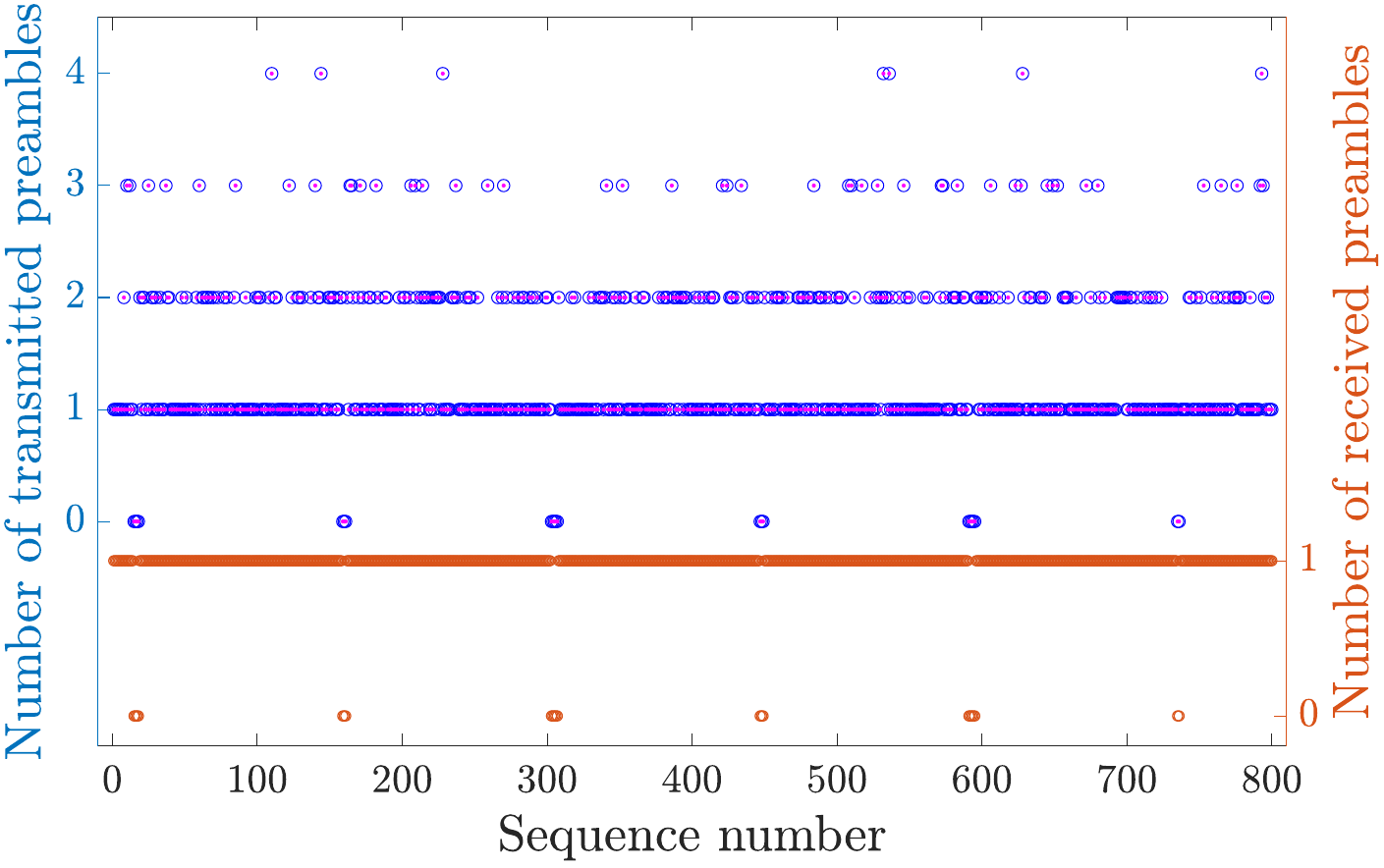}\label{fig:jamming_reference}} \hspace{5mm}
 \subfloat[800 intervals of 60s with active jammer using S1.]{\includegraphics[width=0.95\columnwidth]{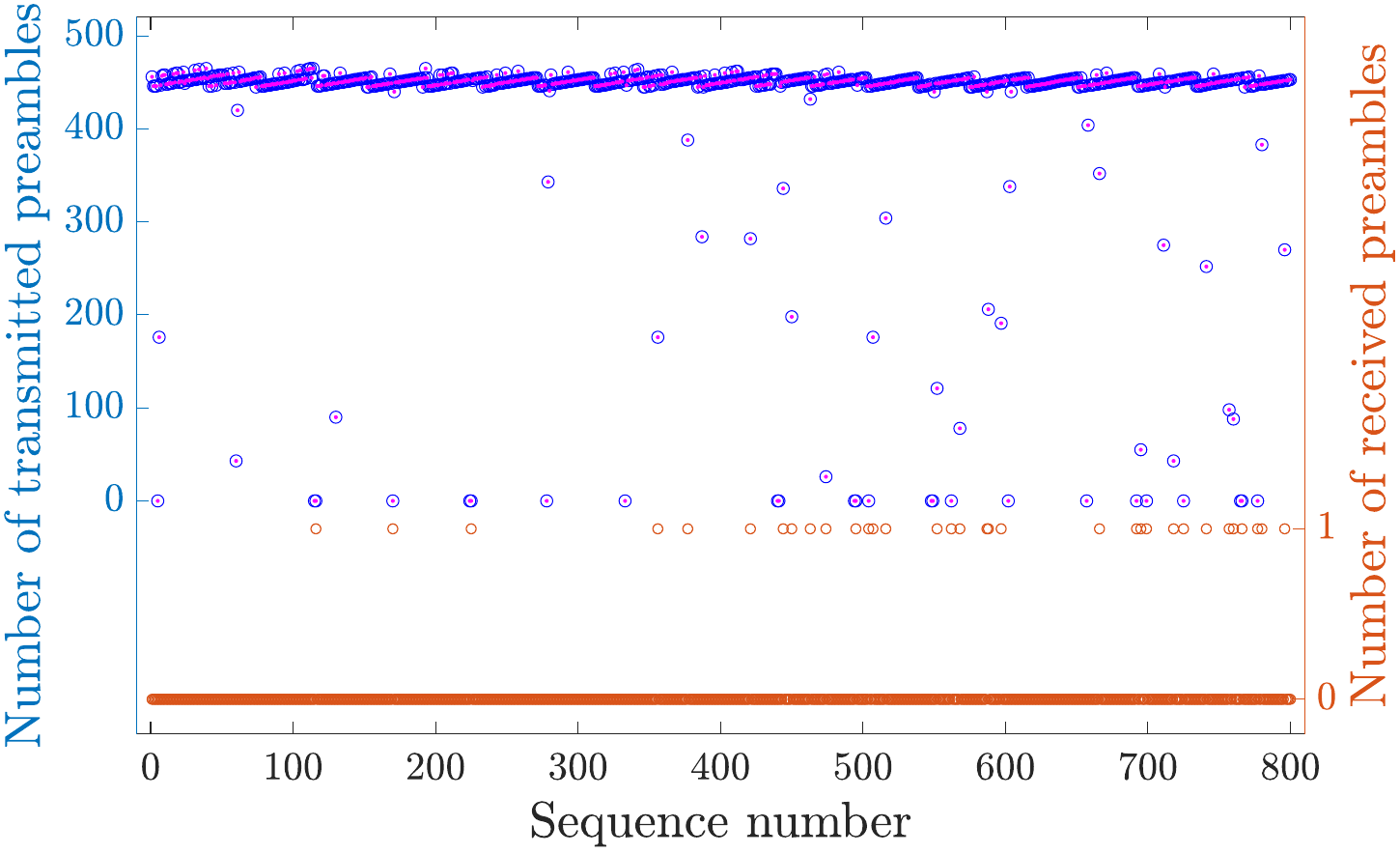}\label{fig:jamming_V1}}
 
\subfloat[800 intervals of 60s with active jammer using S2.]{\includegraphics[width=0.95\columnwidth]{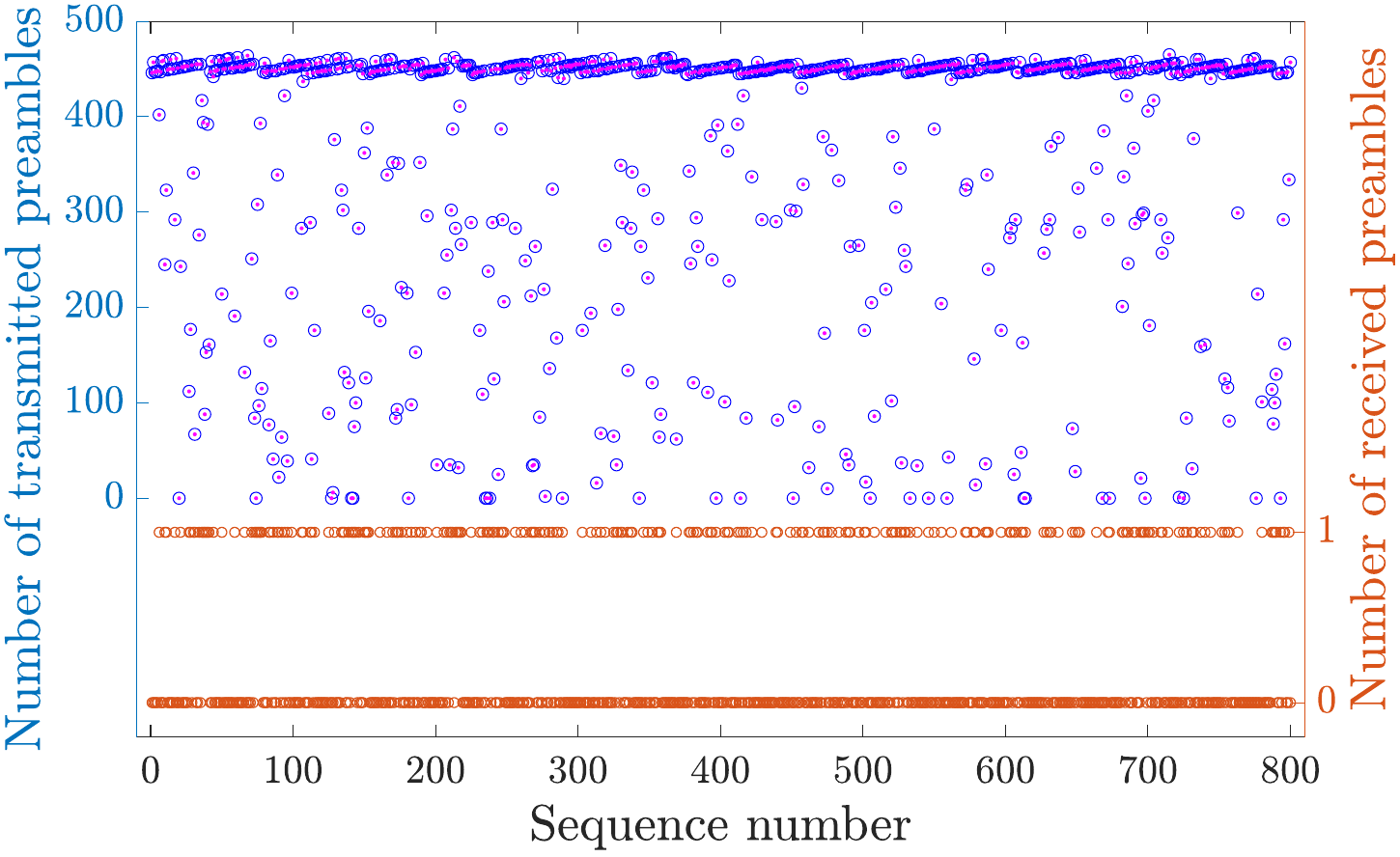}\label{fig:jamming_V2}}\hspace{5mm}
\subfloat[40 intervals of 600s with active jammer using S1.]{\includegraphics[width=0.95\columnwidth]{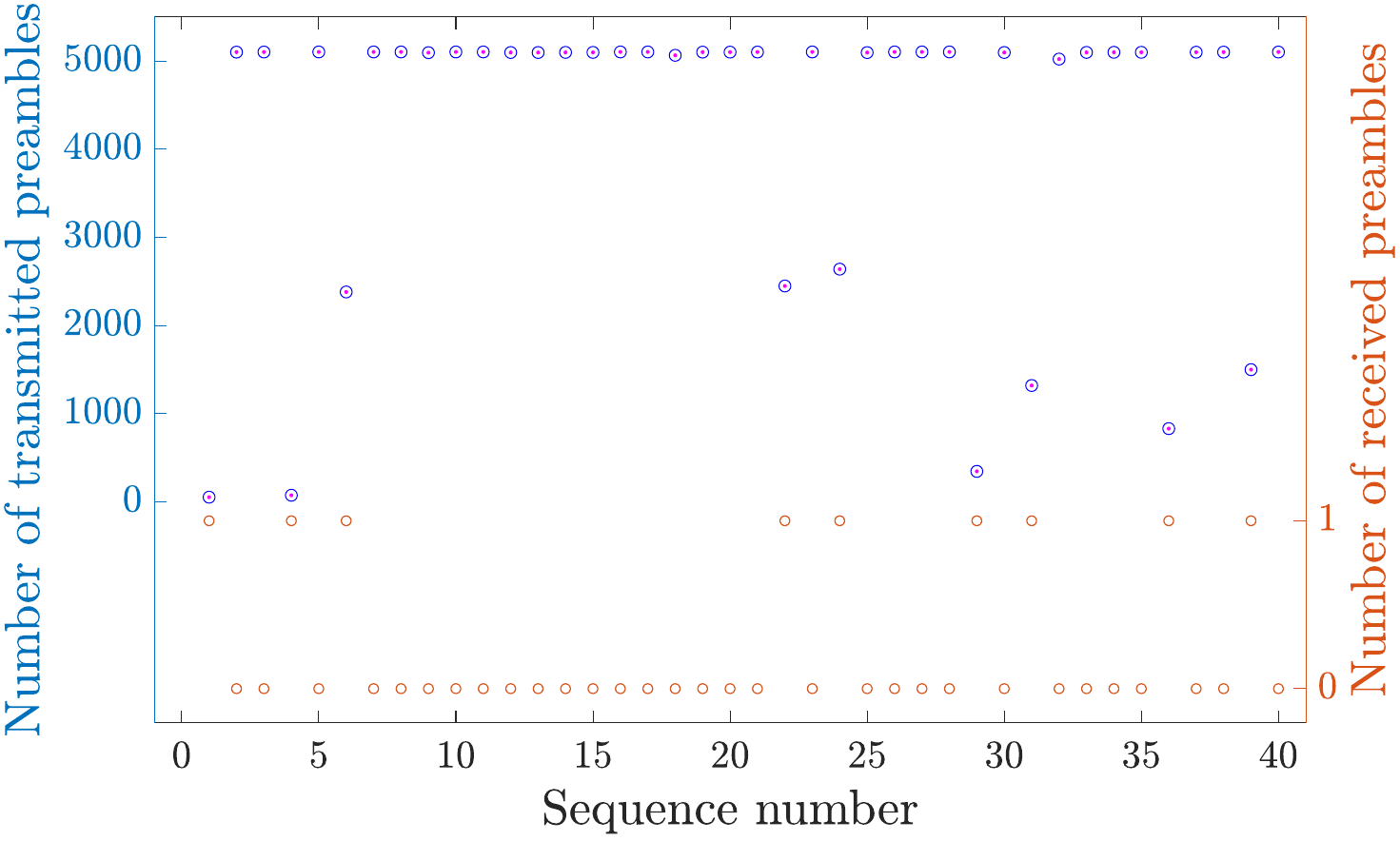}\label{fig:jamming_V1_600}}
  \caption{Number of preambles transmitted by the \gls{uequ} and received by the \gls{gnodb}.}
  \label{fig:results}
\end{figure*}

Figure \ref{fig:jamming_reference} shows the reference measurement series with deactivated jammer, whereas 22 of the total 800 runs were invalid. 
 In the valid 778 runs, a preamble was successfully received from the \gls{gnodb} every time. 
The average number of preambles to be transmitted by the \gls{uequ} for this purpose is 1.39. 
In many cases, the first preamble transmission was already successful. 
Also a successful \gls{ra} procedure took place in any valid interval. 

Next, 800 intervals each were recorded for 60s \gls{uequ} runtime with active jammer for both spectra S1 (Figure \ref{fig:jamming_V1}) and S2 (Figure \ref{fig:jamming_V2}). 
In direct comparison, S1 has significantly better interference properties. As a result, of 787 valid intervals, preamble reception at the \gls{gnodb} was only possible in 33 cases. In total, over 340k preambles were successfully jammed in the entire measurement series. 
This number is lower for S2, at slightly less than 300k. 
Here  the \gls{ra} procedure succeeded in 28.4\%, i.e., in 224 of the 789 valid intervals. 
Since no further preamble is transmitted after the \gls{rar}, the average number of preambles sent in an interval with 374 is also lower compared to S1.



Figure \ref{fig:jamming_V1_600} shows the last series of measurements with S1 and 600s interval duration. 
The percentage of intervals with successful preamble transmission is higher compared to the measurement series with 60~s. 
This is caused by the fact that only one preamble has to be received for a successful \gls{ra}. 
Thus, the probability increases the longer the interval is selected.
Here, the sample size is lower
, but almost 170k preambles were successfully jammed. 
Table \ref{tab:results} summarises the results. 
\begin{table}[h]
\footnotesize
    \centering
	\caption{Overview of the results}
	\label{tab:results}
	\begin{tabular}{@{} c c c c c@{}}
		\toprule[1.5pt]
            \textit{Parameter} & Reference & S1 60s & S1 600s & S2 60s \\
		\midrule[0.8pt]
            $N_{RA,u}$ & 22 & 767 & 31 & 556 \\
            $N_{RA,s}$ & 778 & 33 & 9 & 224 \\
            $N_e$  & 22 & 13 & 0 & 11 \\
            $N_{P,j}$ & - & 343 522 & 169 630 & 295 103\\
            $\bar{N}_{P,t}$ & 1.39 & 437 & 4241 & 374 \\
            $E_{p,j}$ & - & 96 \textit{ppm} & 53 \textit{ppm} & 758 \textit{ppm}  \\ 
            $E_s$ & 100\% & 4.2\% & 22.5\% & 28.4\%  \\
            \bottomrule[1.5pt]
	\end{tabular}
\end{table}

\section{Mitigation} 
Proactive countermeasures are required.
For this, methods for jammer detection must first be investigated. 
In particular, the relationship between the actual damage caused and the scope or costs of the countermeasures must be weighed here.
Resources should be made available step by step for detection and later defence in order to influence regular communication as little as possible \cite{siriwardhana_ai_2021}.  
We expect great benefits from the use of suitable AI algorithms for anomaly detection \cite{lam_machine_2020,fernandez-maimo_dynamic_2019,hussain_deep_2021}. 
And possibly also for the reconstruction of preambles from disturbed \gls{prach} slots. 
Beamforming in \gls{mimo} systems could selectively increase the transmit power or suppress the jammer \cite{akhlaghpasand_jamming_2020}. 
In addition, a spatial rerouting of the preamble via \gls{ris} (\cite{liu_reconfigurable_2021,elmossallamy_reconfigurable_2020}) or D2D communication (\cite{ansari_5g_2018,zhang_beyond_2020}) would be conceivable in order to bypass the jammer or to escape its radius of effect. 
The emerging trend towards cell-less campus networks in the field of inital 6G research may also reduce robustness to PRACH jamming \cite{he_cell-free_2021,azari_evolution_2022}. \\
The interfaces for measurement data acquisition, implementation of neural network structures and access to relevant active radio system parts should therefore already be considered and defined in the standardisation of future network generations. Also a more dynamic and secret distribution of \gls{prach} channels in the resource grid needs further investigation. 

\section{Conclusion}
\label{sec:Concl}
This paper introduced the idea of a smart jammer. Furthermore, setting the PRACH as attack vector was motivated. Additionally, both implementation and testbed were detailed. The results demonstrate a viable threat of low-cost smart jamming on sensitive 5G campus networks using operational data. Only 0.01~\% and 0.1~\% of the preambles reached the \gls{gnodb} for spectrum 1 and spectrum 2, correspondingly. 
As a consequence, the \gls{uequ} was only able to successfully perform an \gls{ra} procedure in 4.2~\% of all 60~s intervals. 
In consecutive 40 long-term measurements of 10 minutes duration, this was also only possible in 22.5~\% of the cases. 
Nevertheless, the jammer here only actively has to jam less than one percent of all time and frequency resources of the cell. 
In reality, this would lead to a \gls{uequ} being unable to perform a \gls{ra} procedure for 10 minutes in more than three quarters of the cases. 
Even at low mobility speeds of the \glspl{uequ}, this has serious consequences. Therefore,  first proposals on the mitigation of a smart \gls{prach} jamming attack are also outlined.

\printbibliography%
\pagebreak

%
%
\end{document}

%% file: organization/preamble.tex
\PassOptionsToPackage{usenames,dvipsnames,svgnames,x11names,table,prologue}{xcolor}%
\PassOptionsToPackage{hyphens}{url}%
\PassOptionsToPackage{nomessages}{fp}%
\ifCLASSINFOpdf
  \usepackage[pdftex]{graphicx}
\else
  \usepackage[dvips]{graphicx}
\fi
\ifCLASSOPTIONcompsoc
   \usepackage[caption=false,font=normalsize,labelfont=sf,textfont=sf]{subfig}
\else
   \usepackage[caption=false,font=footnotesize]{subfig}
\fi

\usepackage{stfloats}
\usepackage[english]{babel}%
\usepackage[utf8]{inputenc}%
\usepackage[babel,style=english]{csquotes}%
\usepackage{hyphsubst}%
\usepackage[%
	activate={true,%
	nocompatibility},%
	final,%
	tracking=true,%
	kerning=true,%
	spacing=true,%
	factor=1100,%
	stretch=10,%
	shrink=10%
]{microtype}%
\usepackage{setspace}%
\usepackage{xcolor}%
\definecolor{todonotecol}{RGB}{250,0,0}%
\usepackage{xparse}
\usepackage[%
	colorlinks=false,%
	urlcolor=black,%
	linkcolor=black,%
	citecolor=black,%
	filecolor=black,%
	breaklinks,%
	]{hyperref}%
\usepackage{url}%
\usepackage[%
	acronym,%
	nopostdot,%
	seeautonumberlist,%
	shortcuts,%
	section=chapter,%
	toc,%
]{glossaries}%
\loadglsentries{./supply/glossaries.tex}
\glsdisablehyper
\usepackage[%
	backend=biber,%
	style=ieee,%
	isbn=false,%
	hyperref=true,%
	maxbibnames=99,%
	sorting=none,%
	natbib=true,%
	language=english,%
	defernumbers=true,%
	]{biblatex}%
\DeclareFieldFormat{sentencecase}{\csname bbx@colon@search\endcsname#1}

\usepackage{nameref}
\usepackage{soul}
\usepackage{flushend}
\usepackage{textcomp}
\usepackage{calc}
\usepackage{xkeyval}
\usepackage{multirow}
\usepackage{tabulary}
\usepackage{makecell}
\usepackage{pgfplots}
\usepackage{tikz}
\usepackage{textcomp} 
\usepackage{verbatim}

%% file: supply/glossaries.tex
\newacronym{ar}{AR}{Application Relation}%
\newacronym{indeco}{InDeCo}{Interconnection Detachment \& Convergence}
\newacronym{coma}{CoMa}{Context-Management}
\newacronym{peac}{PeAc}{Permission \& Access}
\newacronym{dynploy}{DynPloy}{Dynamic Deployment}
\newacronym{remcon}{RemCon}{Remote Configuration}
\newacronym{coagd}{CAD}{Collective Aggregator Discovery}
\newacronym{archdaemon}{ArchDaemon}{Architecture-Daemon}
\newacronym{ctrllog}{ACL}{Aggregated Control Logic}
\newacronym{adagent}{ADAg}{ArchDaemon-Agent}
\newacronym{adaloc}{ADAl}{ArchDaemon-Allocator}
\newacronym{inserv}{InServ}{Inserted-Service}
\newacronym{infrserv}{InfrServ}{Infrastructure-Service}
\newacronym{atserv}{AtServ}{Atop-Service}%
\newacronym{orcus}{OrCuS}{Organic Customization System}
\newacronym{ore}{ORE}{Organic Runtime Environment}%
\newacronym{relag}{Agent}{Agent}%
\newacronym{relorc}{Orc}{Orchestrator}%
\newacronym{slaag}{SLAAg}{Service Level Agreement Agent}%
\newacronym{cmdc}{CMDC}{Centralized Mission and Driving Control}%
\newacronym{dcmod}{DCMod}{Data Collection Module}%
\newacronym{posmod}{PosMod}{Positioning Module}%
\newacronym{mts}{MTS}{Map Transformation Stage}%
\newacronym{sensmod}{SensMod}{Sensing Module}%
\newacronym{sidr}{SIDR}{Sensor ID Resolver}%
\newacronym{camod}{CAMod}{Carrier Aggregation Module}%
\newacronym{muxs}{MuxS}{Multiplexer Stage}%
\newacronym[plural=CSEs,%
	firstplural=Carrier State Estimators (CSEs),%
	longplural=Carrier State Estimators]%
	{cse}{CSE}{Carrier State Estimator}%
\newacronym{ca}{CA}{Carrier Aggregator}%
\newacronym[plural=IAs,%
	firstplural=Information Anchors (IAs),%
	longplural=Information Anchors]%
	{ia}{IA}{Information Anchor}%
\newacronym{dbmod}{DBMod}{Data Base Module}%
\newacronym{bui}{BUI}{Base User Interface}%
\newacronym{bman}{BMan}{Base Manager}%
\newacronym[plural=DSs,%
	firstplural=Data Suppliers (DSs),%
	longplural=Data Suppliers]%
	{ds}{DS}{Data Supplier}%
\newacronym[plural=GWs,%
	firstplural=Gateways (GWs),%
	longplural=Gateways]%
	{gw}{GW}{Gateway}%
\newacronym{auditf}{TTF}{Trust \& Traceability Function}%
\newacronym{specal}{SASF}{Spectrum Allocation \& Sharing Function}%
\newacronym{example}{e.g.}{exempli gratia}
\newacronym{qos}{QoS}{Quality-of-Service}%
\glsunset{qos}%
\newacronym{utc}{UTC}{Coordinated Universal Time}%
\newacronym[plural=KPIs,%
	firstplural=Key Performance Indicators (KPIs),%
	longplural=Key Performance Indicators]%
	{kpi}{KPI}{Key Performance Indicator}%
\newacronym{poc}{PoC}{Proof-of-Concept}%
\newacronym{wip}{WiP}{Work in Progress}%
\newacronym{inft}{IT}{Information Technology}%
\newacronym{fsm}{FSM}{Finite-State Machine}%
\newacronym{fsa}{FSA}{Finite-State Automaton}%
\newacronym[longplural={Distributed Ledger Technologies}]{dlt}{DLT}{Distributed Ledger Technology}%
\newacronym{gps}{GPS}{Global Positioning System}%
\newacronym{trl}{TRL}{Technology Readiness Level}%
\newacronym{opsys}{OS}{Operating System}%
\newacronym{digt}{DT}{Digital Twin}%
\newacronym{rand}{R\&D}{Research \& Development}%
\newacronym{softdr}{SDR}{Software Defined Radio}%
\newacronym{api}{API}{Application Programming Interface}%
\newacronym{uri}{URI}{Uniform Resource Identifier}%
\newacronym{netfunc}{NF}{Network Function}%
\newacronym{nfv}{NFV}{Network Function Virtualization}%
\newacronym{typlv}{TLV}{Type-Length-Value}%
\newacronym{loc}{LoC}{Lines of Code}%
\newacronym{url}{URL}{Uniform Resource Locator}%
\newacronym{nic}{NIC}{Network Interface Card}%
\newacronym{rest}{REST}{Representational State Transfer}%
\newacronym{ftp}{FTP}{File Transfer Protocol}%
\newacronym{http}{HTTP}{Hypertext Transfer Protocol}%
\newacronym{https}{HTTPS}{Hypertext Transfer Protocol Secure}%
\newacronym{rpc}{RPC}{Remote Procedure Call}%
\newacronym{lldp}{LLDP}{Link Layer Discovery Protocol}%
\newacronym{snmp}{SNMP}{Simple Network Management Protocol}%
\newacronym{nsh}{NSH}{Network Service Header}%
\newacronym{sdn}{SDN}{Software-defined Networking}%
\newacronym[plural=SDNs,%
	firstplural=Software-defined Networks (SDNs),%
	longplural=Software-defined Networks]%
	{sdnet}{SDN}{Software-defined Network}%
\newacronym{ip}{IP}{Internet Protocol}%
\glsunset{ip}%
\newacronym{tcp}{TCP}{Transmission Control Protocol}%
\newacronym{udp}{UDP}{User Datagram Protocol}%
\newacronym{dns}{DNS}{Domain Name System}%
\newacronym{ntp}{NTP}{Network Time Protocol}%
\newacronym{ptp}{PTP}{Precision Time Protocol}%
\newacronym{p2p}{P2P}{Point-to-Point}%
\newacronym{peer2p}{Peer2P}{Peer-to-Peer}%
\newacronym{m2m}{M2M}{Machine-to-Machine}%
\newacronym{rbs}{RBS}{Reference Broadcast Time Synchronization}%
\newacronym{rbis}{RBIS}{Reference Broadcast Infrastructure Synchronization}%
\newacronym{mdns}{mDNS}{Multicast DNS}%
\newacronym{dnssd}{DNS-SD}{DNS Service Discovery}%
\newacronym{servoa}{SOA}{Service-Oriented Architecture}%
\newacronym{rtt}{RTT}{Round Trip Time}%
\newacronym{csp}{CSP}{Communication Service Provider}%
\newacronym{mmtc}{MMTC}{Massive Machine Type Communication}%
\newacronym{arq}{ARQ}{Automatic Repeat Request}%
\newacronym{ppm}{PPM}{Parallel Process Migration}%
\newacronym{checkres}{C/R}{Checkpoint/Restore}%
\newacronym{snr}{SNR}{Signal to Noise Ratio}%
\newacronym{acp}{AP}{Access-Point}%
\newacronym{urllc}{URLLC}{Ultra Reliable Low Latency Communication}%
\newacronym{bnetza}{BNetzA}{Bundesnetzagentur}%
\newacronym{5gsba}{SBA}{5G Service Based Architecture}%
\newacronym{sba}{SBA}{Service Based Architecture}%
\newacronym{mano}{MANO}{Management \& Orchestration}%
\newacronym{ran}{RAN}{Radio Access Network}%
\newacronym{cran}{C-RAN}{Centralized - Radio Access Network}%
\newacronym{oran}{O-RAN}{Open - Radio Access Network}%
\newacronym{sbi}{SBI}{Service Based Interface}%
\newacronym{jcas}{JC\&S}{Joint Communication and Sensing}%
\newacronym{rrm}{RRM}{Radio Resource Management}%
\newacronym{uequ}{UE}{User Equipment}%
\newacronym{gmlc}{GMLC}{Gateway Mobile Location Center}%
\newacronym{embb}{eMBB}{enhanced Mobile Broadband}%
\newacronym{mnoper}{MNO}{Mobile Network Operator}%
\newacronym{coretc}{CtC}{Core-to-Core}%
\newacronym{gnodb}{gNB}{gNodeB}%
\newacronym{3gpp}{3GPP}{3rd Generation Partnership Project}%
\newacronym{uwb}{UWB}{Ultra Wideband}%
\newacronym{newr}{NR}{New Radio}%
\newacronym{runit}{RU}{Radio Unit}%
\newacronym{cunit}{CU}{Centralized Unit}%
\newacronym{dunit}{DU}{Distributed Unit}%
\newacronym{cpl}{CP}{Control Plane}%
\newacronym{upl}{UP}{User Plane}%
\newacronym{ngap}{NGAP}{Next Generation Application Protocol}%
\newacronym{amf}{AMF}{Access and Mobility Management Function}%
\newacronym{lmf}{LMF}{Localization Management Function}%
\newacronym{nef}{NEF}{Network Exposure Function}%
\newacronym{nrf}{NRF}{Network Repository Function}%
\newacronym{scp}{SCP}{Service Communication Proxy}%
\newacronym{i40}{I4.0}{Industry 4.0}%
\newacronym{plc}{PLC}{Programmable Logic Controller}%
\newacronym{vpc}{VPC}{Virtual Process Controller}%
\newacronym{tsn}{TSN}{Time-Sensitive Networking}%
\newacronym{cnc}{CNC}{Central Network Controller}%
\newacronym{cuc}{CUC}{Central User Configuration}%
\newacronym{tsncnc}{TSN CNC}{TSN Central Network Controller}%
\newacronym{tssdn}{TSSDN}{Time-Sensitive Software-Defined Networking}%
\newacronym{cpci}{CPCI}{Central Process Control Instance}%
\newacronym{optech}{OT}{Operation Technology}%
\newacronym{iot}{IoT}{Internet of Things}%
\newacronym{iiot}{IIoT}{Industrial Internet of Things}%
\newacronym{agv}{AGV}{Automated Guided Vehicle}%
\newacronym{amr}{AMR}{Autonomous Mobile Robot}%
\newacronym{uav}{UAV}{Unmanned Aerial Vehicle}%
\newacronym{erp}{ERP}{Enterprise Resource Planning}%
\newacronym{ips}{IPS}{Indoor Positioning System}%
\newacronym{vmc}{VMC}{Vehicle Motion Control}%
\newacronym{imu}{IMU}{Inertial Measurement Unit}%
\newacronym{ros}{ROS}{Robot Operating System}%
\newacronym{opc}{OPC}{Open Platform Communication}%
\newacronym{opcua}{\gls{opc} UA}{OPC Unified Architecture}%
\newacronym{mqtt}{MQTT}{Message Queue Telemetry Transport}%
\newacronym{etl}{ETL}{Extract, Transfer, Load}%
\newacronym{dds}{DDS}{Data Distribution Service}%
\newacronym{dgps}{DGPS}{Differential Global Positioning System}%
\newacronym{ecef}{ECEF}{Earth-centered, Earth-fixed}%
\newacronym{gnss}{GNSS}{Global Navigation Satellite System}%
\newacronym{arti}{AI}{Artificial Intelligence}%
\newacronym{mlearn}{ML}{Machine Learning}%
\newacronym{svm}{SVM}{Support Vector Machine}%
\newacronym{sla}{SLA}{Service Level Agreement}%
\newacronym{b2b}{B2B}{Business-To-Business}%
\newacronym{b2c}{B2C}{Business-To-Consumer}%
\newacronym{pmse}{PMSE}{Programme Making and Special Events}%
\newacronym{ppdr}{PPDR}{Public Protection and Disaster Relief}%
\newacronym{cps}{CPS}{Cyber-Physical System}
\newacronym{prach}{PRACH}{Physical Random Access Channel}
\newacronym{pbch}{PBCH}{Physical Broadcast Channel}
\newacronym{pdcch}{PDCCH}{Physical Downlink Control Channel}
\newacronym{pucch}{PUCCH}{Physical Uplink Control Channel}
\newacronym{pdsch}{PDSCH}{Physical Downlink Shared Channel}
\newacronym{pusch}{PUSCH}{Physical Uplink Shared Channel}
\newacronym{fr}{FR}{Frequency Range}
\newacronym{oai}{OAI}{Open Air Interface}
\newacronym{tdd}{TDD}{Time Division Duplexing}
\newacronym{cots}{COTS}{Commercial Off-The-Shelf}
\newacronym{ber}{BER}{Bit Error Rate}
\newacronym{ra}{RA}{Random Access}
\newacronym{stal}{SA}{Stand Alone}
\newacronym{rrc}{RRC}{Radio Ressource Control}
\newacronym{rar}{RAR}{Random Access Response}
\newacronym{cpr}{CP}{Cyclic Prefix}
\newacronym{rel}{RE}{Resource Element}
\newacronym{prb}{PRB}{Physical Resource Block}
\newacronym{sfn}{SFN}{System Frame Number}
\newacronym{mnc}{MNC}{Mobile Network Code}
\newacronym{mcc}{MCC}{Mobile Country Code}
\newacronym{mib}{MIB}{Master Information Block}
\newacronym{sib}{SIB}{System Information Block}
\newacronym{ofdm}{OFDM}{Orthogonal Frequency-Division Multiplexing}
\newacronym{phy}{PHY}{Physical}
\newacronym{fft}{FFT}{Fast Fourier Transformation}
\newacronym{ifft}{IFFT}{Inverse Fast Fourier Transformation}
\newacronym{soc}{SoC}{System-on-a-Chip}
\newacronym{siso}{SISO}{Single Input Single Output}
\newacronym{mimo}{MIMO}{Multiple Input Multiple Output}
\newacronym{usrp}{USRP}{Universal Software Radio Peripheral}
\newacronym{dos}{DoS}{Denial of Service}
\newacronym{ris}{RIS}{Reconfigurable Intelligent Surfaces}
\newacronym{pps}{PPS}{Pulse Per Second}
\newacronym{zc}{ZC}{\textit{Zadoff-Chu}}

%% file: organization/makros.tex
\newcommand{\mytilde}{{\raise.17ex\hbox{$\scriptstyle\mathtt{\sim}$}}}

%% file: organization/settings.tex
\newlength\textheighttemp%
\newlength\textwidthtemp%
\newlength\textheightstd%
\newlength\textwidthstd%
\newlength\textheightold%
\newlength\textwidthold%
\newlength\tempheight%
\newlength\tempwidth%
\SetProtrusion{encoding={*},family={bch},series={*},size={6,7}}
              {1={ ,750},2={ ,500},3={ ,500},4={ ,500},5={ ,500},
               6={ ,500},7={ ,600},8={ ,500},9={ ,500},0={ ,500}}
\SetExtraKerning[unit=space]
    {encoding={*}, family={bch}, series={*}, size={footnotesize,small,normalsize}}
    {\textendash={400,400}, 
     "28={ ,150}, 
     "29={150, }, 
     \textquotedblleft={ ,150}, 
     \textquotedblright={150, }} 
\SetTracking{encoding={*}, shape=sc}{40}

%% file: organization/IEEE_authors-long.tex
\author{%
\IEEEauthorblockN{%
    Julius R. Stegmann\IEEEauthorrefmark{1}, 
    Michael Gundall\IEEEauthorrefmark{1},  %
    and Hans D. Schotten\IEEEauthorrefmark{1}\IEEEauthorrefmark{3} %
    \\%
}%
\IEEEauthorblockA{%
    \IEEEauthorrefmark{1}German Research Center for Artificial Intelligence GmbH (DFKI), Kaiserslautern, Germany \\%
    \IEEEauthorrefmark{3}Department of Electrical and Computer Engineering,  RPTU Kaiserslautern-Landau, Kaiserslautern, Germany %
	\\%
    Email: %
       \{julius\_raphael.stegmann, michael.gundall, hans\_dieter.schotten%
       \}@dfki.de
}%
}%